\documentclass[12pt]{iopart}
\usepackage{graphicx}
\begin{document}
\title{New Kinds of Acoustic Solitons}
\author{S V Sazonov$^1$ and N V Ustinov$^2$}
\address{$^1$ Russian Research Center ''Kurchatov Institute'', 1 Kurchatov sq., 
Moscow 123182, Russia}
\address{$^2$ Tomsk State University, 36 Lenin ave., Tomsk 634050, Russia}
\eads{\mailto{barab@newmail.ru}, \mailto{n\_ustinov@mail.ru}}
\begin{abstract}
We find that the modified sine--Gordon equation belonging to the class of the 
soliton equations describes the propagation of extremely short transverse 
acoustic pulses through the low-temperature crystal containing paramagnetic 
impurities with effective spin $S=\frac12$ in the Voigt geometry case. 
The features of nonlinear dynamics of strain field and effective spins, which 
correspond to the different kinds of acoustic solitons, are studied. 
\end{abstract}
\pacs{05.45.Yv, 43.35.+d, 02.30.Ik}
\vspace{2pc}

The development of physical acoustics has led to the appearance of technical 
tools of producing and measuring acoustic pulses about $10$--$10^2\,\rm ps$ in 
duration \cite{NO,HM}. 
The characteristics of such pulses are very perspective for diagnostics of 
fast processes and spectroscopy of solids. 
This attracts large attention to theoretical study of the interaction of 
picosecond acoustic pulses with paramagnetic crystals and other nonlinear 
media [3--8]. 
Usually, the semiclassical approach is employed to derive the equations 
governing the evolution of acoustic pulses. 
Some of these equations occur to be integrable with the help of the inverse 
scattering transformation (IST) method \cite{ZMNP,N}. 
In particular, the systems of integrable equations that generalize well-known 
integrable models of nonlinear coherent optics \cite{MB} describe the 
propagation of transverse-longitudinal picosecond pulses \cite{Z1,Z2}. 

The duration of picosecond acoustic pulses may be comparable with the 
oscillation period of the quantum transitions involved into the interaction. 
Following well-known parallels between the nonlinear phenomena in coherent 
optics and physical acoustics \cite{Sh,TR}, one has to treat acoustic pulses 
in this case as extremely short pulses \cite{MB,Cas}. 
However, it is necessary in so doing to take into account essential difference 
between acoustic and optical waves. 
The linear velocities of the components of the former can differ significantly 
\cite{Kittel}. 
Thus, the longitudinal component velocity is normally much higher than the 
transverse ones. 
The nonlinear interaction of these components is weak in that case, and, 
consequently, longitudinal and transverse picosecond acoustic pulses propagate 
independently. 
At the same time, transverse components can interact efficiently since 
their linear velocities are equal under propagation along the acoustic 
symmetry axis of the crystal.

In this paper we investigate the nonlinear dynamics of the acoustic extremely 
short pulses in the low-temperature paramagnetic crystal in the external 
magnetic field presence. 
In accordance with above mentioned parallels between coherent optics and 
physical acoustics, we apply here the spectral overlap approximation 
\cite{BN}. 
This approximation is based on condition 
\begin{eqnarray}
\varepsilon\equiv(\omega_0\tau_p)^{2}\ll1,
\label{sc}
\end{eqnarray}
where $\omega_0$ is the characteristic frequency of quantum transitions 
created by the external field; $\tau_p$ is the pulse duration. 
The main aim of the present article is to clarify the role of nonlinear 
interaction of the acoustic pulse components. 
We suppose for this reason that the pulses are especially transverse. 

Let a tetragonal (or cubic) crystal contain paramagnetic impurities with 
effective spin $S=\frac12$. 
Assume that the Cartesian axes $x$, $y$ and $z$ are aligned with symmetry axes 
of the crystal. 
Let the transverse acoustic pulse propagate along the $x$ axis and the 
external magnetic field $\bi{B}$ be parallel to the $z$ axis (Voigt geometry). 
Consider the one-dimensional case with dynamical variables depending on 
coordinate $x$ and time $t$ only. 
Then, the Hamiltonian $\hat H$ of the spin-elastic interaction has the form 
\cite{TR} 
\begin{eqnarray}
\hat H=-\frac{\hbar\omega_0}{2}\left[\hat\sigma_z+F_{44}{\cal E}_{yx}
\hat\sigma_y+F_{55}{\cal E}_{zx}\hat\sigma_z\right]. 
\label{H}
\end{eqnarray}
Here $\omega_0=g\mu_{\rm B}B/\hbar$ is the frequency of the Zeeman splitting 
of the Kramers doublets; $g$ is the Lande factor; $\mu_{\rm B}$ is the Bohr 
magneton; $B=|\bi{B}|$; ${\cal E}_{yx}=\partial u_y/\partial x$ and 
${\cal E}_{zx}=\partial u_z/\partial x$ are the components of the strain 
tensor; $u_y$ and $u_z$ are the Cartesian components of the local displacement 
vector $\bi{u}$; $F_{44}=g^{-1}(\partial g_{yx}/\partial{\cal E}_{yx})_0$ and 
$F_{55}=g^{-1}(\partial g_{zx}/\partial{\cal E}_{zx})_0$ are the components of 
the tensor of the spin-elastic interaction (in Voigt notation; subscript ''0'' 
means differentiation at the absence of acoustic pulse); $g_{jk}$ are the 
components of the Lande tensor; $\hat\sigma_y$ and $\hat\sigma_z$ are the 
Pauli matrices; $\hbar$ is the Planck constant. 
From the microscopic point of view, the spin-elastic coupling appears in the 
case $S=\frac12$ due to the modulation of the Lande tensor components by the 
strain field \cite{TR}. 

In order to achieve fairly efficient interaction between paramagnetic 
impurities and strain field, the Zeeman splitting energy must exceed the 
thermal one.
This implies that paramagnetic crystal has to be at helium temperatures, as it 
was in the experiments on acoustic self-induced transparency \cite{Sh}. 
In that case the self-absorption of hypersound with frequency $10^2\,\rm GHz$ 
(or the picosecond acoustic pulses) due to anharmonicity, defects, etc. is 
appreciably lower than the acoustic absorption due to the presence of 
paramagnetic impurities \cite{TR}. 
Hence, the self-absorption effect playing important role under the room 
temperatures can be ignored in our case. 
Also, characteristic phase relaxation time for transitions within the Zeeman 
multiplets is $10^{-5}$--$10^{-6}\,\rm s$, and the energy relaxation time is 
much longer under such conditions \cite{Sh}. 
We neglect these dissipative effects in what follows because the duration of 
the pulses considered is much shorter than all the relaxation times. 

According to the general scheme of the semiclassical approach, we describe the 
evolution of effective spins by the equation on density matrix $\hat\rho$: 
\begin{eqnarray}
\rmi\hbar\frac{\partial\hat\rho}{\partial t}=[\hat H,\hat\rho].
\label{rho_t}
\end{eqnarray}
On the other hand, the elastic pulse field obeys the classical Hamiltonian 
equation for continuous medium: 
\begin{eqnarray}
\frac{\partial\bi{p}}{\partial t}=-\frac{\delta}{\delta\bi{u}}\left(H_{\rm a}+
\!\int\! n\!<\!\!\hat H\!\!>\!\rmd\bi{r}\right),
\label{Ham_1}
\end{eqnarray}
\begin{eqnarray}
\frac{\partial\bi{u}}{\partial t}=\frac{\delta}{\delta\bi{p}}\left(H_{\rm a}+
\!\int\! n\!<\!\!\hat H\!\!>\!\rmd\bi{r}\right),
\label{Ham_2}
\end{eqnarray}
where $\bi{p}$ is the momentum density of the local displacement of the 
crystal; 
\begin{eqnarray}
H_{\rm a}=\frac12\int\left[\frac{p_y^2+p_z^2}{\rho}+
\rho a^2({\cal E}_{yx}^2+{\cal E}_{zx}^2)\right]\rmd\bi{r}
\label{H_a}
\end{eqnarray}
is the Hamiltonian of the free strain field; $\rho$ is the average density of 
the crystal; $n$ is the concentration of paramagnetic ions; 
$<\!\!\hat H\!\!>\,=\Tr(\hat\rho\hat H)$ is the quantum average value of 
$\hat H$; $a$ is the linear velocity of transverse acoustic waves. 
The integration is carried out over the crystal volume. 

Let us introduce the Bloch variables 
\[
U=\frac{\rho_{21}+\rho_{12}}{2},\qquad V=\frac{\rho_{21}-\rho_{12}}{2\rmi},
\qquad W=\frac{\rho_{22}-\rho_{11}}{2}, 
\]
where $\rho_{jk}$ ($j,k=1,2$) are the elements of the density matrix. 
Then (\ref{rho_t}) gives 
\begin{eqnarray}
\frac{\partial U}{\partial t}=(\omega_0+\Omega_z)V+\Omega_yW,
\label{U_t}
\end{eqnarray}
\begin{eqnarray}
\frac{\partial V}{\partial t}=-(\omega_0+\Omega_z)U,
\label{V_t}
\end{eqnarray}
\begin{eqnarray}
\frac{\partial W}{\partial t}=-\Omega_yU,
\label{W_t}
\end{eqnarray}
where 
\[
\Omega_y=\omega_0F_{44}{\cal E}_{yx},\qquad
\Omega_z=\omega_0F_{55}{\cal E}_{zx}.
\]
With (\ref{H}), (\ref{Ham_1})--(\ref{H_a}) we obtain 
\begin{eqnarray}
\frac{\partial^2\Omega_y}{\partial t^2}-a^2\frac{\partial^2\Omega_y}
{\partial x^2}=-\frac{n\hbar\omega_0^2F_{44}^2}{4\rho}\frac{\partial^2 V}
{\partial x^2},
\label{O_t}
\end{eqnarray}
\begin{eqnarray}
\frac{\partial^2\Omega_z}{\partial t^2}-a^2\frac{\partial^2\Omega_z}
{\partial x^2}=\frac{n\hbar\omega_0^2F_{55}^2}{4\rho}\frac{\partial^2 W}
{\partial x^2}.
\label{O_l}
\end{eqnarray}

Equations (\ref{U_t})--(\ref{O_l}) describe the interaction of the transverse 
strain field with the paramagnetic crystal in the Voigt geometry case. 
As it is seen from (\ref{U_t})--(\ref{W_t}), $y$-component $\Omega_y$ of the 
acoustic pulse causes quantum transitions between the Zeeman sublevels, 
whereas $z$-component $\Omega_z$ shifts dynamically their frequency. 
For transverse acoustic pulse propagating along the $z$ axis (Faraday 
geometry), both components of the pulse excite quantum transitions only. 
The spin-elastic interaction between the components leads in this case to the 
rotation of the polarization plane of the pulse \cite{S} (acoustic Faraday 
effect). 

If we put 
\[
S=W+\rmi U,
\]
then (\ref{U_t}) and (\ref{W_t}) yield 
\begin{eqnarray}
\frac{\partial S}{\partial t}=\rmi(\omega_0+\Omega_z)V+\rmi\Omega_yS. 
\label{S_t}
\end{eqnarray}
Let us assume that $\tau_p\sim10\,\rm ps$ and the orders of $\omega_0$ and 
$\Omega_z$ are comparable. 
Taking $\omega_0\sim10^{10}\,{\rm s}^{-1}$ (that is $B\sim10^3\,\rm Gs$) 
\cite{S,Sh,TR}, we see that condition (\ref{sc}) is valid. 
In that case the first term in the rhs of (\ref{S_t}) can be neglected in the 
approximation of zeroth-order with respect to $\varepsilon$ \cite{ST}. 
Then we have 
\[
S=W_0\rme^{\displaystyle \rmi\theta},
\]
or
\begin{eqnarray}
U=W_0\sin\theta,\qquad W=W_0\cos\theta,
\label{UW}
\end{eqnarray}
where
\begin{eqnarray}
\theta=\int_{t_0}^{t}\Omega_y\,\rmd t',
\label{theta}
\end{eqnarray} 
$W_0$ ($|W_0|\le1/2$) is the inversion of population of the spin sublevels in 
the acoustic pulse absence. 
Substitution (\ref{UW}) into (\ref{V_t}) gives 
\begin{eqnarray}
\frac{\partial V}{\partial t}=-W_0(\omega_0+\Omega_z)\sin\theta.
\label{R_t_t}
\end{eqnarray}

To simplify further the equations we deal with, let us carry out some 
numerical estimations. 
Assuming $W\sim U$, $\partial/\partial t\sim1/\tau_p $ we find from 
(\ref{W_t}) that $\Omega_y\sim1/\tau_p$. 
Therefore, the ratio $\eta_y$ of the rhs of equation (\ref{O_t}) to the terms 
in its lhs is estimated as 
$\eta_y\sim\sqrt{\varepsilon}n\hbar\omega_0F_{44}^2/4\rho a^2$. 
The value of similar parameter of (\ref{O_l}) is estimated as 
$\eta_z\sim\sqrt{\varepsilon}n\hbar\omega_0F_{55}^2/4\rho a^2$. 
For paramagnetic ions $\rm Co^{2+}$ in cubic crystal $\rm MgO$ at helium 
temperatures we use the following experimental data \cite{S,TR}: 
$n\sim10^{19}\,\rm cm^{-3}$, $\omega_0\sim10^{10}\,\rm s^{-1}$, 
$\rho\sim1\,\rm g/cm^3$, $a\sim5\cdot10^5\,\rm cm/s$, and 
$F_{44}\sim F_{55}\sim10^3$. 
If $\tau_p\sim10^{-11}\,\rm s$, then $\eta_y\sim\eta_z\sim10^{-2}$. 
Since parameters $\eta_y$ and $\eta_z$ are much less than unity, we shall 
reduce the order of derivatives in (\ref{O_t}) and (\ref{O_l}) with the help 
of the unidirectional propagation approximation \cite{Eil}. 

Having introduced new independent variables $\tau=t-x/a$ and $\zeta=\eta x$, 
where $\eta=\max(\eta_y,\,\eta_z)$, we obtain 
\[
\frac{\partial}{\partial t}=\frac{\partial}{\partial\tau},\qquad
\frac{\partial}{\partial x}=-\frac{1}{a}\frac{\partial}{\partial\tau}+ 
\eta\frac{\partial}{\partial\zeta}.
\]
In the first order in $\eta$, we write 
\[
\frac{\partial^2}{\partial x^2}\approx\frac{1}{a^2}
\frac{\partial^2}{\partial\tau^2}-2\frac{\eta}{a}
\frac{\partial^2}{\partial\tau\partial\zeta}, 
\qquad
\frac{\partial^2}{\partial x^2}\approx\frac{1}{a^2}
\frac{\partial^2}{\partial\tau^2}
\]
for the lhs and rhs of equations (\ref{O_t}) and (\ref{O_l}), respectively. 
Integration of the wave equations obtained in this way with respect to $\tau$, 
substitution of expressions (\ref{UW}) and taking into account 
(\ref{R_t_t}) give us the following system in the terms of variables $\tau$ 
and $x$: 
\begin{eqnarray}
\frac{\partial \Omega_y}{\partial x}=-\beta_y(\omega_0+\Omega_z)\sin\theta,
\label{O_t_z}
\end{eqnarray}
\begin{eqnarray}
\frac{\partial \Omega_z}{\partial x}=\beta_z\Omega_y\sin\theta,
\label{O_l_z}
\end{eqnarray}
where $\beta_y=-W_0n\hbar\omega_0^2F_{44}^2/(8\rho a^3)$, 
$\beta_z=\beta_yF_{55}^2/F_{44}^2$. 

Equations (\ref{O_t_z}) and (\ref{O_l_z}) possess the integral of motion: 
\begin{eqnarray}
\Omega_z^2+2\omega_0\Omega_z+\frac{F_{55}^2}{F_{44}^2}\Omega_y^2=f(\tau), 
\label{int}
\end{eqnarray}
where function $f(\tau)$ is determined by the boundary conditions.
The similar integral was revealed in \cite{Z2}. 
Defining new variables 
\[
\tau'=\int^\tau_0\sqrt{1+f(\tilde\tau)/\omega_0^2\,}\,\rmd\tilde\tau,
\]
\[
\Omega_y'=\frac{\Omega_y}{\sqrt{1+f(\tau)/\omega_0^2\,}},
\]
\[
\Omega_z'=\frac{\omega_0+\Omega_z}{\sqrt{1+f(\tau)/\omega_0^2\,}}-\omega_0,
\]
one can prove that $f(\tau)$ is supposed equal to zero without loss of the 
generality \cite{Z2}. 
Then, we find from (\ref{int}): 
\begin{eqnarray}
\Omega_z=-\omega_0\left(1-\sqrt{1-\tau_c^2\Omega_y^2}\right),
\label{O_s}
\end{eqnarray}
where 
\[
\tau_c=\frac{F_{55}}{\omega_0F_{44}}. 
\]
(It is seen that inequality $|\Omega_z|\le2\omega_0$ is fulfilled.) 
Finally, using (\ref{theta}), (\ref{O_t_z}) and (\ref{O_s}), we obtain 
\begin{eqnarray}
\frac{\partial^2\theta}{\partial x\partial\tau}=-\omega_0\beta_y
\sqrt{1-\tau_c^2\left(\frac{\partial\theta}{\partial\tau}\right)^2\,}\,
\sin\theta.
\label{msg}
\end{eqnarray}

This equation is reduced to the famous sine--Gordon (SG) equation 
\cite{ZMNP,N} if $\tau_c=0$. 
Equation (\ref{msg}) with $\tau_c\ne0$ is known as the modified SG (mSG) 
equation [19--22] 
and belongs to the class of equations integrable by the IST method. 
Its first physical application was found recently in \cite{SU3}, where 
(\ref{msg}) was shown to describe the propagation of electromagnetic extremely 
short pulses through the anisotropic media. 
In [19--22], 
this equation  was derived in the course of mathematical study of the 
B\"acklund transformation of the SG equation. 

Being integrable with the help of the IST method, (\ref{msg}) admits the zero 
curvature representation 
\begin{eqnarray}
\frac{\partial\hat L}{\partial x}-\frac{\partial\hat A}{\partial\tau}+
[\hat L,\hat A]=0, 
\label{cc}
\end{eqnarray}
where matrices $\hat L$ and $\hat A$ are defined as given 
\[
\hat L=\frac{1}{2\lambda}\left(
\begin{array}{cc}
\rmi\lambda \Omega_y&\sqrt{1-\tau_c^2\Omega_y^2\,}-\rmi\tau_c\Omega_y\\
\sqrt{1-\tau_c^2\Omega_y^2\,}+\rmi\tau_c\Omega_y&-\rmi\lambda \Omega_y
\end{array}
\right),
\]
\[
\hat A=-\frac{\omega_0\beta_y}{2}\left(
\begin{array}{cc}
-\rmi\tau_c\sin\theta&\lambda{\rm e}^{\displaystyle \rmi\theta}\\
\lambda{\rm e}^{\displaystyle-\rmi\theta}&\rmi\tau_c\sin\theta
\end{array}
\right),
\]
and $\lambda$ is the spectral parameter. 
Equation (\ref{cc}) is nothing but the compatibility condition of the 
following Lax pair 
\begin{eqnarray}
\left\{
\begin{array}{l}
\displaystyle{\frac{\partial\xi}{\partial\tau}}_{\mathstrut}=\hat L\xi,\\ 
\displaystyle{\frac{\partial\xi}{\partial x}}^{\mathstrut}=\hat A\xi,
\end{array}
\right.
\label{Lax}
\end{eqnarray}
where $\xi=\xi(\lambda,\tau,x)=(\xi_1,\xi_2)^T$. 

To investigate the nonlinear dynamic of the transverse strain field components 
and effective spins, we construct the soliton solutions of (\ref{msg}). 
It is well known that the multi-soliton solutions of the integrable equation 
can be found using the algebraic methods. 
Here we apply the Darboux transformation (DT) technique \cite{MaSa}. 
Let $\varphi=(\varphi_1,\varphi_2)^T$ be a solution of (\ref{Lax}) with 
$\lambda=\tau_p$. 
The Lax pair (\ref{Lax}) is covariant with respect to DT 
$\{\xi_1,\xi_2,\theta\}\to\{\tilde\xi_1,\tilde\xi_2,\tilde\theta\}$ of the form 
\begin{eqnarray}
\tilde\xi_1=(\lambda\xi_1-\tau_p\varphi_1\xi_2/\varphi_2)
\exp[\rmi(\tilde\theta-\theta)/2],\nonumber\\
\tilde\xi_2=(\lambda\xi_2-\tau_p\varphi_2\xi_1/\varphi_1)
\exp[\rmi(\theta-\tilde\theta)/2],\nonumber\\
\tilde\theta=\theta+\rmi\ln\frac{\tau_p\varphi_1^2-\tau_c\varphi_1\varphi_2}
{\tau_p\varphi_2^2-\tau_c\varphi_1\varphi_2}.
\label{DT1_psi}
\end{eqnarray}
This implies that relation (\ref{DT1_psi}) gives us new solution 
$\tilde\theta$ of the mSG equation (\ref{msg}) if $\theta$ is its known 
solution and $\varphi$ is a solution of the Lax pair. 

In the zero background case (i.e., $\theta=0$), we obtain from (\ref{DT1_psi}) 
the following expression for the one-soliton solution of the mSG equation: 
\[
\theta=2\arccos\frac{q-\tanh\chi}{\sqrt{\Delta}},
\]
where $q=\tau_c/\tau_p$, $\chi=(t-x/v)/\tau_p$, $\Delta=1-2q\tanh\chi+q^2$. 
Velocity $v$ of the soliton and its free parameter $\tau_p$ defining the 
duration are connected by the relation 
\[
v^{-1}=a^{-1}+\omega_0\beta_y\tau_p^2. 
\]
The corresponding formula for $y$-component of the transverse strain field is 
\begin{eqnarray}
\Omega_y=\frac{2\,\,\mbox{sech}\,\chi}{\tau_p}\,\frac{1-q\tanh\chi}{\Delta}. 
\label{O_y_1}
\end{eqnarray}
For ''time area'' $A_y\equiv\int_{-\infty}^{\infty}\Omega_y\,\rmd t$ of this 
component of the acoustic pulse we find 
\[
A_y=\cases{
\pm2\pi&for $|\tau_p|>\tau_c$\\
0&for $|\tau_p|<\tau_c$\\}.
\]

The last formula indicates that the acoustic extremely short pulses are 
divided into two families. 
The family with $A_y=\pm2\pi$ exists for the SG equation also and corresponds 
to unipolar $2\pi$-pulses (kinks and antikinks). 
The pulses of the family with $A_y=0$ are bipolar $0\pi$-pulses. 
Unlike the breathers of the SG equation (for them $A_y=0$ as well), these 
pulses are steady-state. 
The solitons of this kind were called as neutral kinks in \cite{SU3}. 

In the cubic crystal, one has $|F_{55}|=|F_{44}|$. 
Then $|\tau_p|<\tau_c$ due to condition (\ref{sc}), and neutral kinks exist 
only in such a crystal. 
In the crystals with tetragonal symmetry, both types of the solitons are 
possible. 

Let us discuss in details the properties of the acoustic solitons. 
In the case $|\tau_p|>\sqrt2\tau_c$, component $\Omega_y$ (\ref{O_y_1}) of the 
unipolar one-soliton solution has a single maximum, whose value is smaller 
than $1/\tau_c$. 
Accompanying dynamics of effective spins is very similar to that of the SG 
equation: the leading edge of the pulse of $y$-component inverts 
completely the populations of the Zeeman sublevels, and the trailing one 
returns them to the initial state. 
The $z$-component is small as compared to $\omega_0$, and its role is 
insignificant. 

Under $\tau_c<|\tau_p|<\sqrt2\tau_c$, component $\Omega_y$ has two symmetric 
peaks (see solid line in figure~1a) 
\begin{figure}[ht]
\centering
\includegraphics[angle=0,width=2.7in]{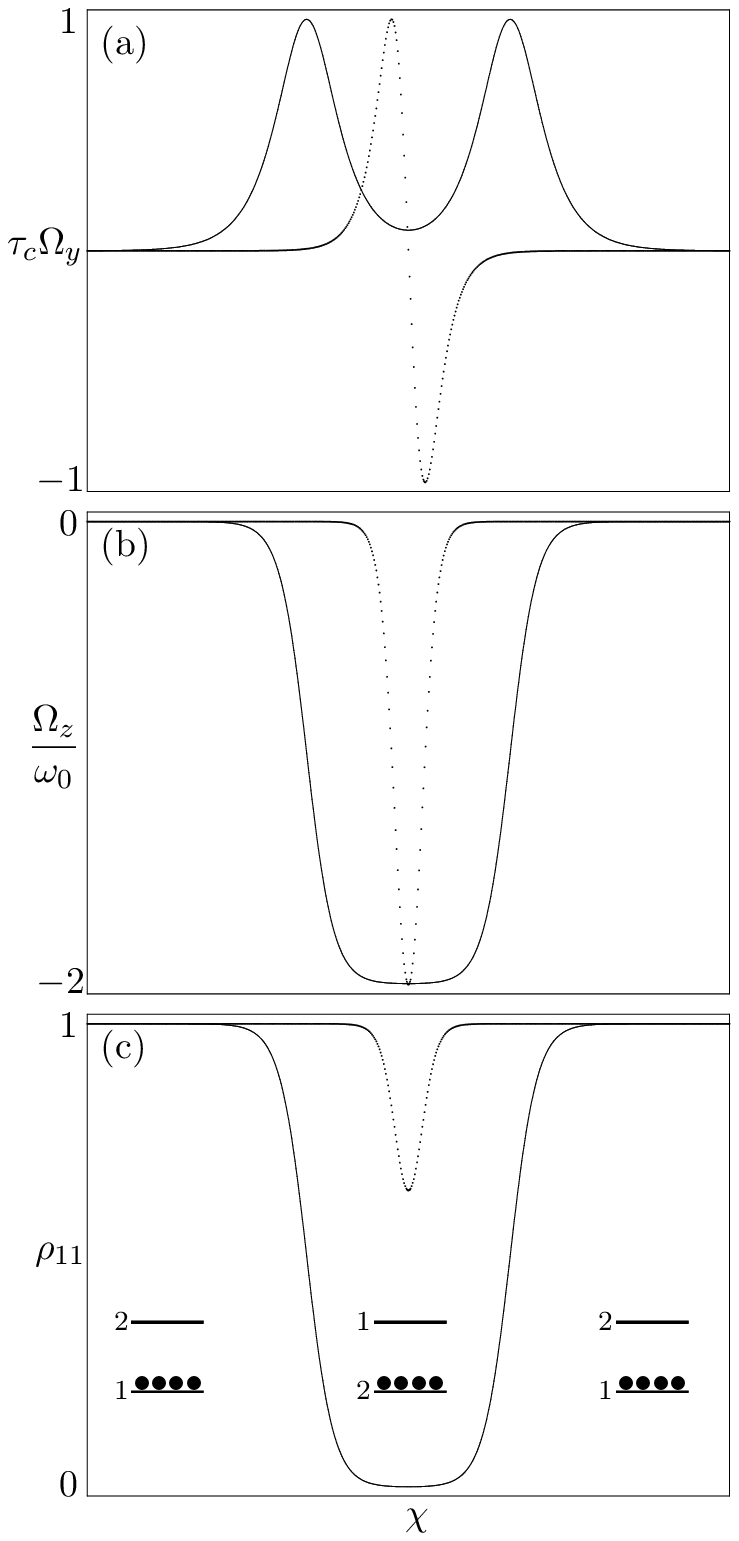}
\caption{Profiles of the components of the strain field and the population 
$\rho_{11}$ for the one-soliton pulses with $\tau_c<|\tau_p|<\sqrt2\tau_c$ 
(solid lines) and $|\tau_p|<\tau_c$ (dotted lines)} 
\end{figure}
with the largest possible amplitude $1/\tau_c$ determined by (\ref{int}). 
The peaks are separated by the time interval 
\[
2|\tau_p|\,\mbox{arcsinh}\sqrt{\frac{q^2}{1-q^2}-1}. 
\]
The first peak inverts the populations of the spin sublevels while the 
component $\Omega_z$ grows in amplitude and reaches the absolute value 
$2\omega_0$ in the center of the soliton (figures~1b and 1c). 
The component $\Omega_z$ has an asymmetry on polarity: it decreases the 
transition frequency ($\Omega_z<0$) and shifts the Zeeman sublevels so that 
the ground sublevel becomes excited. 
On account of this, the paramagnetic ions are in the ground state between the 
peaks. 
(The positions of the spin sublevels $1$ and $2$ of the Kramers doublet under 
the pulse passage are pictured in figure~1c.) 
When the second peak has come, the $z$-component vanishes reverting the mutual 
position of the sublevels to the initial state. 
Finally, the second peak of $y$-component causes the back transitions from 
excited sublevel to the ground one. 

When $|\tau_p|<\tau_c$ and the interval between the peaks of neutral kink 
surpasses its duration, the dynamics of the strain fields and effective spins
is similar to the second case described above. 
The only difference is that the peaks of $\Omega_y$ are opposite in sign. 
The time interval between the peaks is 
\[
2|\tau_p|\,\mbox{arccosh}\sqrt{1+\frac{q^2}{q^2-1}}. 
\]
If we take duration of such a soliton to be shorter, then the peaks are 
brought closer together and the degree of excitation of the paramagnetic ions 
decreases (see dotted lines in figure~1). 

When $|\tau_p|\to\tau_c$, the interval between the peaks grows indefinitely 
large, and $y$-component (\ref{O_y_1}) consists of a single peak with 
amplitude equal to $1/\tau_c$ and with absolute value of time area $A_y$ equal 
to $\pi$. 
This case stresses especially the role of the component $\Omega_z$ of the 
acoustic pulses considered. 
The peak of $\Omega_y$ inverts almost completely the population of the spin 
sublevels. 
This state of the effective spins is unstable in the absence of the strain 
field. 
But, the $z$-component, whose amplitude tends to $2\omega_0$, shifts the 
levels of the Kramers doublets in a such manner that the energy of the excited 
sublevel becomes lesser than the energy of the ground one. 
Owing to this, the state of effective spins after the passage of the 
$y$-component peak becomes stable. 

The form of the acoustic solitons in the case $f(\tau)\ne0$ (see integral 
(\ref{int})) tends to their form in the case $f(\tau)=0$ at $t\to\infty$ since 
the pulse velocity $v$ differs from linear velocity $a$ of the transverse 
waves. 
As it follows from the previous consideration, this means in particular that 
the amplitude of the pulses is bounded, and the component of the strain field 
parallel to the external magnetic field has the asymmetry on polarity. 

In this paper we considered the propagation of the transverse acoustic 
extremely short pulse through paramagnetic crystal in a direction 
perpendicular to external magnetic field. 
It was shown that the dynamics of the strain field and effective spins is 
governed by the modified sine--Gordon equation (\ref{msg}). 
The soliton solutions of this equation reveal strong nonlinear coupling 
between the components of the acoustic pulse. 
As a result of this, the behaviour of paramagnetic impurities and elastic 
fields during the interaction exhibits new features. 

\section*{Acknowledgment}
This work is supported by the Russian Foundation for Basic Research (Grant 
\#\,05--02--16422).


\section*{References}
\begin{thebibliography}{99}
\bibitem{NO} Naugolnykh K and Ostrovsky L 1998 {\it Nonlinear Wave Processes 
in Acoustic} (Cambridge: Cambridge University Press)
\bibitem{HM} Hao H-Y and Maris H J 2001 \PR B {\bf64} 064302 
\bibitem{A} Adamashvili G T 1999 {\it Physica} B {\bf266} 173 
\bibitem{S} Sazonov S V 2000 {\it JETP} {\bf91} 16 
\bibitem{VS} Voronkov S V and Sazonov S V 2001 {\it JETP} {\bf93} 236 
\bibitem{Z1} Zabolotskii A A 2003 {\it JETP} {\bf96} 1089 
\bibitem{Z2} Zabolotskii A A 2003 \PR E {\bf67} 066606 
\bibitem{SU1} Sazonov S V and Ustinov N V 2006 {\it JETP} {\bf102} 741 
\bibitem{ZMNP} Zakharov V E, Manakov S V, Novikov~S~P and Pitaevskii~L~P 1984
{\it Theory of Solitons: The Inverse Scattering Method} (New York: Consultants 
Bureau) 
\bibitem{N} Newell A C 1985 {\it Solitons in Mathematics and Physics} 
(Philadelphia: SIAM) 
\bibitem{MB} Maimistov A I and Basharov A M 1999 {\it Nonlinear Optical Waves}
(Dortrecht: Kluwer Acad. Publ) 
\bibitem{Sh} Shiren N S 1970 \PR B {\bf2} 2471 
\bibitem{TR} Tucker J W and Rampton V W 1972 {\it Microwave Ultrasonics in 
Solid State Physics} (Amsterdam: North--Holland) 
\bibitem{Cas} Casperson L W 1998 \PR A {\bf57} 609 
\bibitem{Kittel} Kittel C 1976 {\it Introduction to Solid States Physics} 5th 
ed. (New York: Wiley) 
\bibitem{BN} Belenov \'E M and Nazarkin A V 1990 {\it JETP Lett.} {\bf51} 288 
\nonum Belenov \'E M, Nazarkin A V and Ushchapovskii~V~A 1991 {\it Sov. Phys. 
JETP} {\bf73} 422 
\bibitem{ST} Sazonov S V and Trifonov E V 1994 \jpb {\bf27} L7 
\bibitem{Eil} Eilbeck J C 1972 \JPA {\bf5} 1355 
\nonum Eilbeck J C, Gibbon J D, Caudrey~P~J and Bullough~R~K 1973 \JPA
{\bf6} 1337 
\bibitem{K} Kruskal M D 1974 {\it Lect. Appl. Math.} {\bf15} 61 
\bibitem{Ch} Chen H-H 1974 \PRL {\bf33} 925 
\bibitem{Nak} Nakamura A 1980 \JPSJ {\bf49} 1167 
\bibitem{BZ} Borisov A B and Zykov S A 1998 {\it Theor. Math. Phys.} {\bf115} 
530 
\bibitem{SU3} Sazonov S V and N V Ustinov 2006 {\it JETP Letters} {\bf83} 483 
\nonum Sazonov S V and N V Ustinov 2006 {\it JETP} {\bf103} 561 
\bibitem{MaSa} Matveev V B and Salle M A 1991 {\it Darboux transformations and 
solitons} (Berlin--Heidelberg: Springer--Verlag) 
\end{thebibliography}
\end{document}